\documentclass[12pt]{article}
 \usepackage{epsfig}
 \def\be{\begin{equation}}
 \def\ee{\end{equation}}
 \def\bea{\begin{eqnarray}}
 \def\eea{\end{eqnarray}}
 \usepackage{graphicx}

 \catcode`\@=11
 \def\lsim{\mathrel{\mathpalette\@versim<}}
 \def\gsim{\mathrel{\mathpalette\@versim>}}
 \def\@versim#1#2{\vcenter{\offinterlineskip
 \ialign{$\m@th#1\hfil##\hfil$\crcr#2\crcr\sim\crcr } }}
 \catcode`\@=12

 \catcode`@=12
\begin{document}
 \thispagestyle{empty}
 \begin{flushright}
 UCRHEP-T598\\
 May 2019\
 \end{flushright}
 \vspace{0.6in}
 \begin{center}
 {\LARGE \bf Scotogenic Cobimaximal Dirac Neutrino\\ Mixing 
from $\Delta(27)$ and $U(1)_\chi$\\}
 \vspace{1.2in}
 {\bf Ernest Ma\\}
 \vspace{0.2in}

{\sl Physics and Astronomy Department,\\ 
University of California, Riverside, California 92521, USA\\}
\vspace{0.1in}
\end{center}
 \vspace{1.2in}

\begin{abstract}\
In the context of $SU(3)_C \times SU(2)_L \times U(1)_Y \times U(1)_\chi$, 
where $U(1)_\chi$ comes from $SO(10) \to SU(5) \times U(1)_\chi$, supplemented 
by the non-Abelian discrete $\Delta(27)$ symmetry for three lepton families, 
Dirac neutrino masses and their mixing are radiatively generated through dark 
matter.  The gauge $U(1)_\chi$ symmetry is broken spontaneously.  The 
discrete $\Delta(27)$ symmetry is broken softly and spontaneously.  
Together, they result in two residual symmetries, a global $U(1)_L$ lepton 
number and a dark symmetry, which may be $Z_2$, $Z_3$, or $U(1)_D$ 
depending on what scalar breaks $U(1)_\chi$.  Cobimaximal neutrino mixing, 
i.e. $\theta_{13} \neq 0$, $\theta_{23} = \pi/4$, and 
$\delta_{CP} = \pm \pi/2$, may also be obtained. 

\end{abstract}

\newpage
\baselineskip 24pt

\noindent \underline{\it Introduction}~:~\\
Whereas there exist powerful theoretical arguments that neutrinos are 
Majorana, there is yet no incontrovertible experimental evidence that 
they are so, i.e. no definitive measurement of a nonzero neutrinoless 
double beta decay.  To make a case for neutrinos to be Dirac, 
the existence of a right-handed neutrino $\nu_R$ must be justified, 
which is of course not required in the standard model (SM) of quarks 
and leptons.  The canonical choice is to extend the SM gauge symmetry 
$SU(3)_C \times SU(2)_L \times U(1)_Y$ to the left-right symmetry 
$SU(3)_C \times SU(2)_L \times SU(2)_R \times U(1)_{(B-L)/2}$. 
In that case, the $SU(2)_R$ doublet $(\nu,e)_R$ is required, and the 
charged $W_R^\pm$ gauge boson is predicted along with a neutral $Z'$ gauge 
boson.

A more recent choice is to consider $U(1)_\chi$ which comes from 
$SO(10) \to SU(5) \times U(1)_\chi$, with $SU(5)$ breaking to the SM 
at the same grand unified scale. 
Assuming that $U(1)_\chi$ survives to an intermediate scale, the current 
experimental bound on the mass of $Z_\chi$ being about 
4.1 TeV~\cite{atlas-chi-17,cms-chi-18}, then $\nu_R$ 
must exist for the cancellation of gauge anomalies.  Now $\nu_R$ is a 
singlet and $W_R^\pm$ is not predicted.  In this context, new insights into 
dark matter~\cite{m18, m19-1} and Dirac neutrino masses~\cite{m19-2,m19-3} 
have emerged.  

To make sure that $\nu_R$ itself does not have a Majorana mass, the breaking 
of $U(1)_\chi$ should not come from a scalar which couples to $\nu_R \nu_R$. 
This simple idea was first discussed~\cite{mpr13} in 2013 in the case of 
singlet fermions charged under a gauge $U(1)_X$.  If the latter is broken 
by a scalar with three units of $X$ charge, it is impossible for these 
fermions with one unit of $X$ charge to acquire Majorana masses.  Hence 
the residual symmetry is global U(1) in this case.  It is  
straightforward then to apply this idea to lepton number~\cite{ms15}.  

A second issue regarding Dirac neutrinos is that the corresponding 
Yukawa couplings linking $\nu_L$ to $\nu_R$ through the 
SM Higgs boson must be very small.  To avoid these tree-level 
couplings, it is often assumed that some additional symmetry exists 
which forbids these dimension-four couplings, but Dirac neutrino masses 
may be generated radiatively as this symmetry is broken softly by 
dimension-three terms.  For a generic discussion, see Ref.~\cite{mp17}, 
which is patterned after that for Majorana neutrinos~\cite{m98}.  
In some applcations~\cite{gs08,fm12,bmpv16}, the particles in the loop 
belong to the dark sector.  This is called the scotogenic mechanism, 
from the Greek 'scotos' meaning darkness, the original one-loop 
example~\cite{m06} of which was applied to Majorana neutrinos.

Instead of the {\it ad hoc} extra symmetry which forbids the tree-level 
couplings, exotic assignments of the gauge charges of $\nu_R$ may be 
used~\cite{ms15,yd18,bccps18,cryz18,dkp19} instead.  However, a much more 
efficacious idea is to use a non-Abelian discrete family symmetry, which is  
softly broken in the dark sector.  In this paper, 
$\Delta(27)$~\cite{m06-1,vkr07,m13-1,abmpv14} is shown to be useful in 
achieving the goal of having scotogenic Dirac neutrino masses with a mixing 
pattern~\cite{bmv03,gl04,mn12} called 
cobimaximal~\cite{m15-1,h15,m16-1,m16-2,fgjl16,gl17}, 
i.e. $\theta_{23} = \pi/4$ and $\delta_{CP} = \pm \pi/2$, which is consistent 
with present neutrino oscillation data~\cite{t2k18} for $\delta_{CP} = -\pi/2$.

\noindent \underline{\it Outline of Model}~:~\\
The particles of this model are shown in Table 1. 
\begin{table}[tbh]
\centering
\begin{tabular}{|c|c|c|c|c|c|c|c|c|c|c|}
\hline
particle & $SO(10)$ & $SU(3)_C$ & $SU(2)_L$ & $U(1)_Y$ & 
$U(1)_\chi$ & $\Delta(27)$ & $U(1)_L$ & $Z_2^D$ & $Z_3^D$ & $U(1)_D$ \\
\hline
$(\nu,e)$ & 16 & 1 & 2 & $-1/2$ & 3 & 3 & 1 & + & $1$ & 0 \\ 
$e^c$ & 16 & 1 & 1 & 1 & $-1$ & $1_1,1_7,1_4$ & $-1$ & + & $1$ & 0 \\ 
$\nu^c$ & 16 & 1 & 1 & 0 & $-5$ & $3^*$ & $-1$ & $+$ & $1$ & 0 \\ 
\hline
$N$ & $126^*$ & 1 & 1 & 0 & 10 & $3$ & 1 & $-$ & $\omega^2$ & $-1$ \\ 
$N^c$ & 126 & 1 & 1 & 0 & $-10$ & $3^*$ & $-1$ & $-$ & $\omega$ & 1 \\ 
\hline
\hline
$(\phi_1^0,\phi_1^-)$ & 10 & 1 & 2 & $-1/2$ & $-2$ & $3^*$ & 0 & + & 1 & 0 \\ 
$(\phi_2^+,\phi_2^0)$ & 10 & 1 & 2 & $1/2$ & 2 & 3 & 0 & + & 1 & 0 \\ 
\hline
$(\eta^+,\eta^0)$ & 144 & 1 & 2 & 1/2 & 7 & 1 & 0 & $-$ & $\omega^2$ & $-1$ \\ 
$\sigma$ & 16 & $1$ & 1 & $0$ & $-5$ & 1 & $0$ & $-$ & $\omega$ & 1 \\ 
\hline
\hline 
$\zeta_2$ & 126 & 1 & 1 & 0 & $-10$ & 1 & 0 & + &  &  \\
\hline
$\zeta_3$ & 672 & 1 & 1 & 0 & 15 & 1 & 0 &  & 1 &  \\
\hline
$\zeta_4$ & 2772 & 1 & 1 & 0 & $-20$ & 1 & 0 &  &  & 0 \\
\hline
\end{tabular}
\caption{Particle content of model.}
\end{table}

In the notation above, all fermion fields are left-handed.  The usual 
right-handed fields are denoted by their charge conjugates.  The SM 
particles transform under $U(1)_\chi$ according to their $SO(10)$ origin, 
as well as the particles of the dark sector $(N, N^c, \eta, \sigma)$. 
The input family symmetry is $\Delta(27)$.  The gauge $U(1)_\chi$ is broken 
by $\zeta_2$ or $\zeta_3$ or $\zeta_4$.  In each case, a residual $U(1)_L$ 
symmetry remains for lepton number whereas the dark symmetry becomes 
$Z_2^D$ or $Z_3^D$ or $U(1)_D$ respectively. 
The complete Lagrangian is invariant under gauge $U(1)_\chi$ in all 
its terms, as well as $\Delta(27)$ in all the dimension-four terms. 
Whereas the breaking of gauge $U(1)_\chi$ must only be spontaneous,  
through the vacuum expectation value of $\zeta_2$ or $\zeta_3$ or $\zeta_4$, 
the breaking of $\Delta(27)$ is both spontaneous, through the vacuum 
expectation values of $\Phi_{1,2}$, and explicit, through soft 
dimension-three terms as shown below. 

The key feature of this model is the interplay between $U(1)_\chi$ and 
$\Delta(27)$ for restricting the interaction terms among the various 
fermions and scalars.  The irreducible representations of $\Delta(27)$ 
and their character table are given in Ref.~\cite{m06-1}.  The important 
point is that if a set of 3 complex fields transforms as the 3 
representation of $\Delta(27)$, then its conjugate transforms as $3^*$, 
which is distinct from 3.  The basic multiplication rules are
\begin{equation}
3 \times 3 = 3^* + 3^* + 3^*, ~~~ 3 \times 3^* = \sum^9_{i=1} 1_i.
\end{equation}
From Table 1, the Yukawa term $e e^c \phi_1^0$ is allowed, but not 
$\nu \nu^c \phi_2^0$ because of $\Delta(27)$.  Furthermore, the usual 
dimension-five operator for Majorana neutrino mass, i.e. 
$\nu \nu \phi_2^0 \phi_2^0$, is forbidden as well as the usual singlet 
Majorana mass term $\nu^c \nu^c$.  Note that without $U(1)_\chi$, 
$\nu^c \nu^c$ is a soft term breaking $\Delta(27)$ and would then have been  
allowed by itself.  To obtain Dirac neutrino masses, the scalar doublet 
$\eta$ and singlet $\sigma$ with odd $Q_\chi$ as well as the fermion 
singlets $N,N^c$ with even $Q_\chi$ are added.  Note that they belong 
to the dark sector because SM fermions have odd $Q_\chi$ and the SM 
Higgs doublet has even $Q_\chi$, as pointed out in Ref.~\cite{m18}. 

Since $\Phi_1^\dagger$ transforms 
exactly like $\Phi_2$, the linear combination 
$\Phi = (v_1 \Phi_1^\dagger + v_2 \Phi_2)/\sqrt{v_1^2+v_2^2}$ 
is the analog of the standard-model Higgs doublet, where 
$\langle \phi^0_{1,2} \rangle = v_{1,2}$.  In the following, only $\Phi$ 
is used, and because it is a 3 under $\Delta(27)$, it is denoted 
as $\Phi_{1,2,3}$.
The dark scalars and fermions have allowed interactions with $\nu,\nu^c$ 
and $\Phi$ under $U(1)_\chi$.  The soft breaking of $\Delta(27)$ then 
allows the one-loop generation of radiative Dirac neutrino masses as shown 
in Fig.~1.
\begin{figure}[htb]
\vspace*{-5cm}
\hspace*{-3cm}
\includegraphics[scale=1.0]{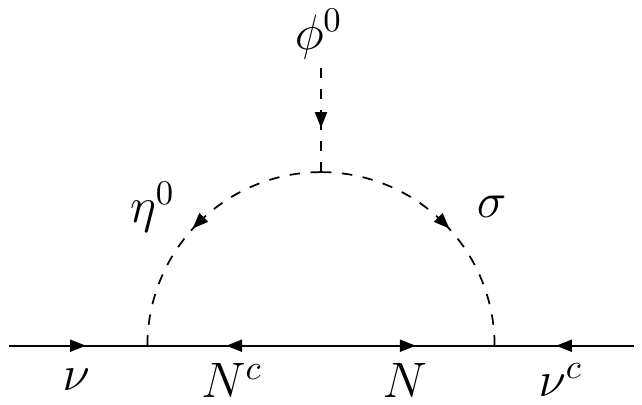}
\vspace*{-21.5cm}
\caption{One-loop diagram for scotogenic $U(1)_\chi$ Dirac neutrino 
mass.}
\end{figure}
The dimension-four terms, i.e. $\nu N^c \eta^0$, $\nu^c N \sigma$, respect 
both $U(1)_\chi$ and $\Delta(27)$.  The dimension-three scalar trilinear 
couplings $\eta^0 \sigma \bar{\phi}^0$ respect $U(1)_\chi$ but not 
$\Delta(27)$.  The dimension-three $N^c N$ terms are also allowed to 
break $\Delta(27)$.

Consider now the spontaneous breaking of $U(1)_\chi$.  First, because 
$\nu, N \sim 3$ and $\nu^c, N^c \sim 3^*$ under $\Delta(27)$, they 
cannot obtain Majorana masses through any scalar which is a singlet. 
Hence $U(1)_L$ lepton number holds as indicated in Table 1.  If 
$\zeta_2$ is used, then the term $\zeta_2^* \sigma^2$ is allowed, hence 
the residual dark symmetry is $Z_2^D$.  Similarly, $\zeta_3 \sigma^3$ 
yields $Z_3^D$ with $\omega = \exp(2\pi i/3)$, and $\zeta_4$ yields 
$U(1)_D$ because $\zeta_4^* \sigma^4$ 
is not allowed by renormalizability, and cannot be generated by the given 
particle content of the model.
In the case of $Z_3^D$ from $\zeta_3$, there is an 
equivalent assignment of lepton number and dark symmetry.  Instead 
of the conventional thinking that $N,N^c$ must carry lepton number, 
the latter may be assigned~\cite{m15} to the scalars $\eta$ and $\sigma$. 
Hence lepton number becomes $Z_3^L$~\cite{mpsz15} with 
$\nu, \sigma \sim \omega$; $\nu^c, \eta \sim \omega^2$; 
$N, N^c, \phi \sim 1$; whereas dark symmetry remains $Z_3^D$ with 
$\sigma, N^c \sim \omega$; $\eta, N \sim \omega^2$; $\nu, \nu^c, \phi \sim 1$.

\noindent \underline{\it Cobimaximal Neutrino Mixing}~:~\\
Using the decomposition $3 \times 3^*$ and $\langle \phi^0_i \rangle = v_i$, 
with $1_1,1_7,1_4$ as defined in Ref.~\cite{m06-1}, instead of the usual 
$1_1,1_2,1_3$ of the original $A_4$ model~\cite{mr01} of neutrino mixing,  
the charged-lepton mass matrix is given by
\begin{equation}
{\cal M}_l = \pmatrix{f_e v_1^* & f_\mu v_3^* & f_\tau v_2^* \cr 
f_e v_2^* & f_\mu v_1^* & f_\tau v_3^* \cr 
f_e v_3^* & f_\mu v_2^* & f_\tau  v_1^*} = 
\pmatrix{m_e & 0 & 0 \cr 0 & m_\mu & 0 \cr 0 & 0 & m_\tau},
\end{equation}
where $v_2=v_3=0$ has been assumed for the spontaneous breaking of 
$\phi^0_{1,2,3}$.  This ${\cal M}_l$ is diagonal and different from 
that of Ref.~\cite{mr01}.  
It allows also three independent masses for the charged leptons, and 
the emergence of lepton flavor triality~\cite{m10,cdmw11} in the 
Yukawa interactions of the three charged leptons with the three 
Higgs doublets.

In the neutrino sector, the tree-level Yukawa couplings $\nu \nu^c \phi^0$ 
are forbidden by $\Delta(27)$.  Hence the $3 \times 3$ Dirac neutrino 
mass matrix ${\cal M}_\nu$ is generated through dark matter (scotogenic) 
as shown in Fig.~1.  Since $\eta^0 \sigma \bar{\phi}^0$ is just one coupling, 
the flavor structure of ${\cal M}_\nu$ comes from the $N^cN$ mass terms 
which break $\Delta(27)$ softly.  Assuming the residual symmetry to be 
generalized $N_2 - N_3$, $N^c_2 - N^c_3$, and $N_{1,2,3} - N^c_{1,3,2}$ 
exchange with complex conjugation~\cite{gl04}, 
the $3 \times 3$ $N^cN$ mass matrix is of the form
\begin{equation}
{\cal M}_N = \pmatrix{A & D & D^* \cr D & B & C \cr D^* & C & B^*},
\end{equation}
where $A,C$ are real.  The above is exactly of the form~\cite{bmv03,gl04,mn12} 
required for cobimaximal 
mixing~\cite{m15-1,h15,m16-1,m16-2,fgjl16,gl17}, i.e. 
$\theta_{13} \neq 0$, $\theta_{23} = \pi/4$, and $\delta_{CP} = \pm \pi/2$, 
because the neutrino basis is also the one where the charged leptons 
are diagonal.  If $B_I=D_I=0$, then $\theta_{13}=0$ and $\theta_{23} = \pi/4$. 
If 
\begin{equation}
A-B_R-C+D_R=0
\end{equation}
in addition, then $\tan^2 \theta_{12} = 1/2$ as well, 
i.e. tribimaximal mixing is obtained.  Since neutrino oscillation data 
are close to this limit, the above quantities may be considered small 
if not zero, hence the neutrino mass eigenvalues are approximately 
given by
\begin{equation}
m_{\nu_1} \simeq (2A + B_R + C - 4D_R)/3, ~~ 
m_{\nu_2} \simeq (A + 2B_R + 2C + 4D_R)/3, ~~ 
m_{\nu_3} \simeq B_R - C.
\end{equation}
If $B_I \neq 0$ or $D_I \neq 0$ or both, cobimaximal mixing is obtained. 
However, $\theta_{13}$ and $\theta_{12}$ are not fixed. If Eq.~(4) is 
valid together with $B_I = 2D_I$, then~\cite{m12-1}
\begin{equation}
\tan^2 \theta_{12} = {1 - 3 \sin^2 \theta_{13} \over 2} < {1 \over 2}, 
\end{equation}
in good agreement with data.

\noindent \underline{\it Dark Sector}~:~\\
To compute the Dirac neutrino mass matrix of Fig.~1, assume first that 
$\eta^0 \sigma$ couples only to $\bar{\phi}^0_{1}$, leading to the 
possibility of lepton flavor triality~\cite{m10,cdmw11} which may be tested 
experimentally.  Note then that the one-loop calculation is equivalent 
to taking the difference of the exchanges of two scalar mass eigenstates
\begin{equation}
\chi_1 = \cos \theta ~\sigma - \sin \theta ~\bar{\eta}^0, ~~~ 
\chi_2 = \sin \theta ~\sigma + \cos \theta ~\bar{\eta}^0,  
\end{equation}
where $\theta$ is the mixing angle due to the $\bar{\phi}^0 \eta^0 \sigma$ 
term.  Let the $\nu_i N^c_k \eta^0$ Yukawa coupling be $h_L {U}_{ik}$ and 
the $\nu^c_j N_k \sigma$ Yukawa coupling be $h_R {U}^T_{kj}$, then the 
Dirac neutrino mass matrix is given by
\begin{equation}
({\cal M_\nu})_{ij} = {h_L h_R \sin 2 \theta \over 16 \pi^2} 
\sum_k {U}_{ik}  M_k \left[ {m_2^2 \over m_2^2-M_k^2} 
\ln {m_2^2 \over M_k^2} - {m_1^2 \over m_1^2-M_k^2} 
\ln {m_1^2 \over M_k^2} \right] {U}^T_{kj},
\end{equation}
where $m_{1,2}$ are the masses of $\chi_{1,2}$ and $M_k$ is the mass of 
$N_k$.  If $|m_2^2-m_1^2| << m_2^2 + m_1^2 = 2m_0^2 << M_k^2$, then 
the $M_k$ contribution reduces to
\begin{equation}
M_k~[~~~] =  {(m_2^2-m_1^2) \over M_k} \left[ \ln {M_k^2 \over m_0^2} - 
1 \right].
\end{equation}
This expression is of the radiative seesaw form.  On the other hand, 
if $M_k << m_{1,2}$, then~\cite{m12}
\begin{equation}
M_k~[~~~] = \ln(m_2^2/m_1^2) M_k.
\end{equation}
This is no longer a seesaw formula.  It shows that the three Dirac neutrinos 
$\nu$ may have masses which are proportional to those of the three  
dark Dirac fermions $N$.  

If Eq.~(9) holds, then the lighter of $\chi_{1,2}$ is dark matter.  It 
should not have a large $\eta^0$ component because it would then have 
significant interactions with nuclei through the $Z$ gauge boson and be 
ruled out by underground direct-search experiments.  Assuming $\theta$ in 
Eq.~(7) to be very small, then it should be $\chi_1$.  If its mass is 
greater than that of the SM Higgs boson, its annihilation to the latter 
is a well-known mechanism for generating the correct dark-matter relic 
abundance of the Universe.

If Eq.~(10) holds, then the lightest $N$ is dark matter, i.e. $N_1$ for 
the normal hierarchy of neutrino masses $(m_{\nu_1} < m_{\nu_2} < m_{\nu_3})$ 
or $N_3$ for the inverse hierarchy $(m_{\nu_3} < m_{\nu_1} < m_{\nu_2})$. 
In either case, the other two $N$s will decay to the lightest $N$ plus 
a neutrino pair or charged lepton pair, through $\chi_{1,2}$ or $\eta^\pm$. 
The annihilation of $N \bar{N} \to \nu \bar{\nu}$ through $\chi_1$ 
exchange has a cross section $\times$ relative velocity given by
\begin{equation}
\sigma \times v_{rel} = {h_R^4 \over 32 \pi} {M_N^2 \over (m_1^2 + M_N^2)^2},
\end{equation}
assuming again that $\theta$ is very small.  As a numerical 
example, let $M_N =150$ GeV, $m_1 = 400$ GeV, $h_R = 0.62$, then this is 
about 1~pb, which is a typical value for obtaining the correct 
dark-matter relic abundance of the Universe, i.e. $\Omega h^2 = 0.12$.

At the mass of 150 GeV, the constraint on the elastic scattering cross 
section of $N$ off nuclei is about $1.5 \times 10^{-46}$~cm$^2$ from 
the latest XENON result~\cite{xenon18}.  This puts a lower limit on the 
mass of $Z_\chi$, i.e. 
\begin{equation}
\sigma_0 = {4m_P^2 \over \pi} {[Z f_P +(A-Z) f_N]^2 \over A^2} 
< 1.5 \times 10^{-10}~{\rm pb},
\end{equation}
where
\begin{equation}
f_P = g^2_{Z_\chi} N_V (2u_V + d_V)/M^2_{Z_\chi}, ~~~ 
f_N = g^2_{Z_\chi} N_V (u_V + 2d_V)/M^2_{Z_\chi}, ~~~ 
\end{equation}
and $Z=54$, $A=131$ for xenon.  In $U(1)_\chi$, the vector couplings are 
\begin{equation}
N_V = \sqrt{5 \over 2}, ~~~ u_V =0, ~~~ d_V = {-1 \over \sqrt{10}}.
\end{equation}
Using $\alpha_\chi = g^2_{Z_\chi}/4\pi = 0.0154$ from Ref.~\cite{m18}, 
the bound $M_{Z_\chi} > 16$ TeV is obtained.  If $M_N = 6$ GeV as 
considered in Ref.~\cite{m19-3}, this bound may be lowered to 4.5 TeV. 

\noindent \underline{\it Concluding Remarks}~:~\\
The $U(1)_\chi$ gauge symmetry and a minimal particle content with a 
softly broken $\Delta(27)$ family symmetry are the ingredients for the 
radiative generation of Dirac neutrino masses through dark matter. 
Both symmetries are broken, but the resulting residual symmetries, 
i.e. global $U(1)_L$ and $Z_2^D$ or $Z_3^D$ or $U(1)_D$, maintain the 
Dirac nature of neutrinos and the stability of dark matter.

The charged-lepton mass matrix is diagonal with the choice of 
$1_1,1_7,1_4$ representations for $e^c_{1,2,3}$, and $v_2=v_3=0$ for 
the Higgs symmetry breaking.  The Dirac neutrino mass matrix comes from 
the soft breaking of $\Delta(27)$ in the $N^cN$ mass matrix in the dark 
sector.  Using generalized exchange symmetries with complex conjugation, 
this is shown to be of the form resulting in cobimaximal mixing, 
i.e. $\theta_{13} \neq 0$, $\theta_{23} = \pi/4$, 
$\delta_{CP} = \pm \pi/2$, in agreement with present data for 
$\delta_{CP} = -\pi/2$.

The dark-matter candidate is either mostly a singlet scalar $\chi_1$ 
or a Dirac fermion $N$, both of which have $Z_\chi$ interactions. 
The bound on $M_{Z_\chi}$ depends on the mass of the dark matter, 
but is about 16 TeV for $M_N = 150$ GeV.

\noindent \underline{\it Acknowledgement}~:~\\
This work was supported in part by the U.~S.~Department of Energy Grant 
No. DE-SC0008541.  

\baselineskip 22pt

\bibliographystyle{unsrt}

\end{document}